\documentstyle[preprint,aps]{revtex}
\begin{document}
\vspace{-80ex}
\baselineskip=24pt
\begin{flushright}

NUHEPTH-95-05\\

\it{June, 1995}
\end{flushright}

\begin{centerline}
{\large\bf The Hyperfine Spin Splittings In Heavy Quarkonia}
\end{centerline}
\bigskip
\begin{center}
Yu-Qi Chen\cite{chen} and Robert J. Oakes\cite{oakes}
\end{center}

\begin{center}
Department of Physics and Astronomy, Northwestern University, Evanston,
IL 60208
\end{center}

\vspace{1cm}
\vspace{-10mm}
\begin{abstract}
The hyperfine spin splittings in heavy quarkonia are studied using
the recently developed renormalization group improved spin-spin
potential which is independent of the scale parameter $\mu$. The
calculated energy difference between the $J/\psi$ and the $\eta_c$
fits the experimental data well, while the predicted energy difference
$\Delta M_p$ between the center of the gravity of $1^3P_{0,1,2}$
states and the $1^1P_1$ state of charmonium has the correct sign but
is somewhat larger than the experimental data. This is not surprising
since there are several other contributions to $\Delta M_p$, which
we discuss, that are of comparable size ($\sim 1$ MeV) that should
be included, before precise agreement with the data can be expected.
The mass differences of the $\psi'-\eta_c'$, $\Upsilon(1S)-\eta_b$,
$\Upsilon(2S)-\eta_b'$, and $B_c^*-B_c$ are also predicted.

\end{abstract}


\newpage
\section{ Introduction}

The hyperfine splittings in heavy quark-antiquark systems can provide
information about  strong interactions or Quantum Chromodynamics (QCD)
at low energies. Since the motion of the heavy quark and  antiquark in
heavy quarkonia is nonrelativistic, their dynamics can be well described
by nonrelativistic potential models. The hyperfine splittings arise from
higher order relativistic corrections and can be calculated  using
perturbation theory, given the appropriate spin-dependent potential.
Recently, significant progress has been made in the theoretical study
of the spin dependent potential\cite{0}. In this work the spin-dependent
potentials were derived from QCD first principles using the Heavy Quark
Effective Theory (HQET)\cite{18}. The spin-dependent potential was
separated into short distance parts involving  Wilson coefficients and
long distance parts which were expressed in terms of  gauge invariant
correlation functions of the color-electric and  color-magnetic fields
weighted by the Wilson loop path integral\cite{1}. If the tree level
values for the Wilson coefficients are used the potential reduces to
Eichten's and Feinberg's result\cite{1}. And using the one-loop values
of the Wilson coefficients, also calculating the correlation functions
to one-loop in perturbation theory, the spin-dependent potential at the
one-loop level in perturbation QCD\cite{4a,4b} is recovered. However,
the leading logarithmic terms appearing in perturbative calculations
were also summed up in Ref.\cite{0} using the Renormalization Group
Equation (RGE) to obtain a scale independent result. Therefore, the
spin-dependent heavy quark-antiquark potential derived  in Ref.\cite{0}
is scale-independent and thus improves upon and generalizes both
Eichten's and Feinberg's result\cite{1} and the one-loop perturbative
result\cite{4a,4b}. In addition, this improved result\cite{0} satisfies
all the general relations among the different parts of the spin-dependent
potential\cite{2}. In the following we use this improved, more general
potential\cite{0} to calculate the hyperfine splittings in the $c\bar{c}$,
$c\bar{b}$, and $b\bar{b}$ systems. Specifically, we calculate the energy
difference between the $^3S_1$ and the $^1S_0$ states and the difference
$\Delta M_p$ between the center of the gravity of the $^3P_{0,1,2}$ states
and the $^1P_1$ state.

First we note that the $P$-wave hyperfine splitting $\Delta M_p$ in
charmonium has been experimentally determined to be $-0.9\pm 0.2$
MeV\cite{7}, which is not only much smaller than the splittings caused
by the spin-orbit and the tensor interactions, but also the $S$-wave
hyperfine splittings, which typically are $50-100$ MeV. Naively, one
might estimate the hyperfine splitting to be smaller than the spin-orbit
and tensor splittings by the order of $v^2$, where $v$ is the relative
quark-antiquark velocity, or about 1/10 in charmonium. This interesting
point has been studied previously\cite{8,9,10,11,13,14}. According to the
Fermi-Breit formula, which follows from lowest order perturbation theory,
the hyperfine spin splitting is proportional to the wavefunction at the
origin, which vanishes for $P$-waves. However, one-loop corrections give
logarithmic terms that are nonlocal and  allow a non-zero contribution
to the $P$-wave hyperfine splittings. Several previous
calculations\cite{8,9,10,11}  of $\Delta M_p$ used only the one-loop
perturbative spin-spin potential\cite{4a,4b} and the results are remarkably
close to the experimental value. This agreement with the experimental
value of $\Delta M_p$, taking into account only the one-loop contribution
is surprising, since there are other contributions to $\Delta M_p$ of
similar size; for example, nonlocal terms coming from higher orders. We
will discuss such effects below, as it is instructive to see how these
higher order contributions could affect the results.

In the following we will use the general formula for the spin-spin part of
the renormalization-group-improved spin-dependent potential that was derived
in Ref.\cite{0} to calculate the hyperfine spin splittings in the $c\bar{c}$,
$b\bar{b}$, and $c\bar{b}$ systems. Since the spin-spin potential is a short
distance feature, perturbation theory can reliably be used in the calculation.
Our result for the $1^3S_1-1^1S_0$ splitting between the $J/\psi$ and the
$\eta_c$  agrees well with the experimental value and our predictions for
the mass differences $\psi'-\eta_c'$, $\Upsilon(1S)-\eta_b$,
$\Upsilon(2S)-\eta_b'$, and $B_c^*-B_c$ are reasonable. However, the
contribution to the $P$-wave energy difference, $\Delta M_p$, between the
center of the gravity of $1^3P_{0,1,2}$ states and the $1^1P_1$ state,
while having the correct sign, is somewhat larger than the experimental
data. That is, when the contributions of the leading logarithmic terms
are summed up and included, the agreement with that data is not as good
as when only the one-loop perturbative spin-spin potential, in which the
leading logarithmic contributions are not summed up and included is used.
We will discuss the implications of these results in greater detail below
and point out that there are several other contributions to the rather
small energy difference $\Delta M_p$ which  estimates indicate are of the
same order of magnitude as the spin-spin contribution. It therefore appears
that the agreement of the one-loop perturbative result with the data is
probably fortuitous.

The following Sec. II is devoted to the calculational methods. In Sec. III
we present our numerical results and in Sec. VI we discuss these results
and our conclusions.

\section{calculational methods}

To calculate the hyperfine splittings in the heavy quark-antiquark
systems we will use the spin-spin part of the renormalization-group
improved general formula for the spin-dependent potential\cite{0}
derived in the framework of HQET\cite{18}. In the derivation the
renormalized two-particle effective Lagrangian was first calculated to
order $1/m^2$. Then, treating the terms of higher order in $1/m$ in the
effective Lagrangian as perturbations, the four point Green's function
on the Wilson loop\cite{24} with the time interval $T$ was calculated
in the limit where $m\to \infty$ first followed by $T\to \infty$\cite{22,1}.
In this limit, using  standard perturbative methods, the large $T$
behavior of the Green's function is of the form
\begin{eqnarray}
I &\propto & e^{-T\epsilon(m,r)}.
\label{en}
\end{eqnarray}
{}From Eq.~(\ref{en}) $\epsilon(m,r)$, the potential energy between the
quark and the antiquark, can be extracted. Expanding $\epsilon(m,r)$ in
powers of $1/m$ each of the spin-dependent potentials can be factorized
into a short distance part, involving Wilson coefficients, and a long
distance part, which can be expressed in terms of correlation functions
of the color-electric and color-magnetic fields weighted by the Wilson-loop
integral. Using the notation of Ref.\cite{0}, the resulting  spin-spin
potential is
\begin{eqnarray}
\Delta H_{ss}(m_1,m_2,r) &=&
\displaystyle\frac{{\rm\bf{S}_1\cdot {S}_2}}{3m_1m_2}
\left[\;c_3(\mu,m_1)c_3(\mu,m_2)V_4(\mu,r)
-6N_c g_s^2(\mu) d(\mu)\delta^3({\rm\bf r})\;\right],
\label{hss}
\end{eqnarray}
where $m_1$, $m_2$, and ${\bf S}_1$, ${\bf S}_2$ are the masses and the
spins of the heavy quark and the antiquark, respectively, $\mu$ is the
renormalization subtraction point, $N_c$ is the number of colors, and
$g_s(\mu)$ is the running coupling constant. The Wilson coefficients
$c_3(\mu,m)$ and  $d(\mu)$ were calculated in leading logarithmic
approximation in Ref.\cite{21} and Ref.\cite{0}, respectively, and are
\begin{eqnarray}
c_3(\mu,m) &=&
\left(\displaystyle\frac{\alpha_s(\mu)}{\alpha_s(m_1)}\right)^{-{9\over 25}},
\label{c3}
\end{eqnarray}
and
\begin{eqnarray}
d(\mu) &=& \displaystyle\frac{N_c^2-1}{8N_c^2}  c_3(m_2,m_1)[1-c_3^2(\mu,m_2)]
\nonumber \\ &=&\displaystyle\frac{N_c^2-1}{8N_c^2}
\left(\displaystyle\frac{\alpha_s(m_2)}{\alpha_s(m_1)}\right)^{-{9\over 25}}
\left[ 1-\left(\displaystyle\frac{\alpha_s(\mu)}{\alpha_s(m_2)}\right)^
{-{18\over 25}} \right].
\label{dmu}
\end{eqnarray}
In Eq.~(\ref{hss}) $V_4(\mu,r)$ is the color magnetic-magnetic correlation
function which can be expressed as
\begin{eqnarray}
V_4(\mu,r) &\equiv &
\lim_{T\to \infty}\displaystyle\int^{T/2}_{-T/2}dz
\int^{T/2}_{-T/2}dz' \frac{g_s^2(\mu) }{T}
\langle B^i({\rm\bf x}_1,z)B^i({\rm\bf x}_2,z')\rangle /\langle
1\rangle ,
\label{v4}
\end{eqnarray}
where $\langle \cdots\rangle$ is defined by
\begin{equation}
\langle \cdots\rangle \equiv \int [dA^\mu]Tr \left\{ P\left[\exp
\left(ig\oint_{C(r,T)} dz_\mu A^\mu(z) \right)
\cdots\right]\right\}_{x\in C} \exp (iS_{YM}(A)),
\label{corr}
\end{equation}
Here $C(r,T)$ represents the Wilson loop\cite{24}, $P$ denotes the
path ordering, and $r\equiv |{\rm\bf x}_1- {\rm\bf x}_2|$.

We emphasize that this is a general result for the hyperfine part of the
spin-dependent potential to order $1/m^2$.  It absorbs the short distance
contributions to the potential into the coefficients $c_3(\mu,m)$ and
$d(\mu)$ while the long distance contributions to the potential are
contained in the correlation function $V_4(\mu,r)$. Moreover, the result
is independent of the factorization scale since the $\mu$-dependence in
the coefficients cancels the $\mu$-dependence in the correlation function.
The first term in the bracket in Eq.~(\ref{hss}) is a nonlocal term while
the second term is a local one which is generated by mixing with the first
(nonlocal) term under renormalization. We note that if the coefficients are
evaluated at tree level; i.e., $c_3(\mu,m)=1$ and $d(\mu)=0$, the potential
reduces to the Eichten-Feinberg result\cite{1}. And if these coefficients
are expanded to order $\alpha_s(\mu)$ and the correlation function is also
evaluated only to one-loop, the logarithmic terms in Eq.~(\ref{hss}) then
reduce to the one-loop spin-spin potential\cite{4a,4b}. Therefore, this
renormalization-group improved potential, Eq.~(2), extends both Eichten's and
Feinberg's result\cite{1} as well as the one-loop perturbative
potential\cite{4a,4b}, containing each of these results as special cases.

In a heavy quarkonium state the typical momentum transfer is of order
$m\,v$, where $v$ is the relative velocity of the heavy quark and the
antiquark, and the typical size is of order $1/(mv)$. In such a low
momentum region the correlation function $V_4(\mu,r)$ could in principle
have nonperturbative contributions and should, therefore, be calculated
using nonperturbative methods. However, since confinement in QCD is
color-electrical, it is reasonable to expect the color-magnetic field to
be predominately a short distance effect. Thus the color magnetic-magnetic
correlation function $V_4(\mu,r)$ should fall off quite fast when the
distance $r$ becomes large. This is confirmed both by lattice
calculations\cite{26} and by the experimental fact that $\Delta M_p$ is
empirically very  small. If $V_4(\mu,r)$ had a significant long distance
component  $\Delta M_p$ would be considerably larger, contrary to the data.
We therefore can safely assume that the potential $V_4(\mu,r)$ is a short
distance effect which can be calculated using perturbative QCD. Its
perturbative expression can be obtained from the following arguments:
As mentioned above, the result for $\Delta H_{ss}(r)$ in Eq.~(\ref{hss})
is $\mu$-independent since the $\mu$-dependence in the coefficients
cancels the $\mu$-dependence in the correlation function. However, to
explicitly demonstrate this cancellation to all orders one must calculate
the correlation function to all orders, which is impossible to do directly.
Fortunately, using the RGE, this can be done in the leading logarithmic
approximation. In momentum space the Fourier transformation of $V_4(\mu,r)$,
denoted by $\widetilde{V}_4(\mu,q)$, is  dimensionless and is only a
function of the two variables $\mu$ and $q$. It must, therefore, be a
function of $\ln(q^2/\mu^2)$ and these logarithms can be summarized  using
the RGE in the effective theory. Alternately, there is another simple
approach: If we choose $\mu=q$ all of these logarithmic terms vanish and
only the tree level term $\displaystyle\frac{N_c^2-1}{N_c}g^2_s(\mu)|_{\mu=q}$
remains in $\widetilde{V}_4(\mu,q)$ in the leading logarithmic approximation.
Then all the nonlocal logarithmic terms having been absorbed into the
coefficients $c_3(q,m)$ and $d(q)$. Consequently, we find the hyperfine part
of the spin-dependent potential in momentum space in the leading logarithmic
approximation to be
\begin{eqnarray}
\Delta \widetilde{H}_{ss}(m_1,m_2,q)  &=&
\displaystyle\frac{{\rm\bf{S}_1\cdot {S}_2}}{3m_1m_2}  g_s^2(q)
\left[\;
\frac{N_c^2-1}{N_c}c_3(q,m_1)c_3(q,m_2) -6N_c d(q)\;\right],
\label{hssq}
\end{eqnarray}
where $c_3(q,m)$ and $d(q)$ are given by Eqs.~(\ref{c3}) and (\ref{dmu}).
This final formula, Eq.~(\ref{hssq}), for the hyperfine spin-dependent
potential, which we will use in our calculations of the hyperfine
spin-splittings in the $c\bar{c}$, $c\bar{b}$, and $b\bar{b}$ systems,
improves upon the one-loop perturbative calculation in two important
respects: ({\it{i}}) it is  independent of $\mu$ and ({\it{ii}}) it includes
the higher order logarithmic terms.

To first order perturbation theory in $\Delta H_{ss}$ the energy shift
caused by $\Delta H_{ss}(r)$ is
\begin{eqnarray}
\Delta E &=& \displaystyle \int d^3{\rm\bf r} \Psi^*_{l,l_z}({\rm\bf r} )
\Delta H_{ss} (r) \Psi_{l,l_z}({\rm\bf r})
\end{eqnarray}
where $\Psi_{l,l_z}({\rm\bf r}) $ is the nonrelativistic wavefunction of
the bound state with total angular momentum $l$ and $z$-component $l_z$.
For simplicity we suppress spin and color indices  and retain only the
space-dependent indices. Separating the radial part, $u(r)$, we write
$\Psi_{l,l_z}({\rm\bf r}) $ as
\begin{eqnarray}
\Psi_{l,l_z}({\rm\bf r}) &=& u(r)\; Y_{l,l_z}(\theta,\phi),
\end{eqnarray}
where $Y_{l,m}(\theta,\phi)$ are the standard spherical harmonics.
Rotational invariance implies that $\Delta E$ is independent of $l_z$.
Averaging over $l_z$  and using properties of the spherical harmonics,
$\Delta E $ can be expressed as
\begin{eqnarray}
\Delta E &=& \displaystyle \int \frac{d^3{\rm\bf r} }{4\pi}
|u(r)|^2 \Delta H_{ss} (r) .
\end{eqnarray}
Taking the Fourier transform, in momentum space $\Delta E$ is given by
\begin{eqnarray}
\Delta E &=& \displaystyle \int \frac{d^3{\rm\bf q} }{(2\pi)^3}\xi(q)
\Delta \widetilde{ H}_{ss} (q) ,
\end{eqnarray}
where
\begin{eqnarray}
\xi(q) &=& \displaystyle \int \frac{d^3{\rm\bf r} }{4\pi}
e^{i{\rm\bf q\cdot r }} |u(r)|^2
= \frac{1}{q} \int dr\,r\sin qr |u(r)|^2 .
\label{xiq}
\end{eqnarray}
Finally, doing the angular integration we have
\begin{eqnarray}
\Delta E &=& \displaystyle \frac{1}{2\pi^2}
\int dq\,q^2 \xi(q) \Delta \widetilde{H}_{ss} (q),
\label{de}
\end{eqnarray}
which we will use to numerically calculate the hyperfine splittings.
Of course, the radial wave function $u(r)$ must first be obtained by
numerically solving the Schr\"{o}dinger equation with some chosen
potential and then $\xi(q)$ [Eq.~(\ref{xiq})] can easily be calculated
by using the Fast Fourier Transformation program.

\section{Numerical results}

Using Eqs.~(\ref{hssq}), (\ref{xiq}), and (\ref{de}) the hyperfine spin
splittings for both the $S$-wave and the $P$-wave states were numerically
calculated. The radial wavefunction was obtained by numerically solving
the Schr\"{o}dinger equation. For comparison, we used three popular
potential models. One was the Cornell model\cite{30} in which the potential
has the form,
\begin{eqnarray}
 V(r) &=& \displaystyle -\frac{\kappa}{r} +\frac{r}{a^2},
\end{eqnarray}
with
\begin{equation}
\begin{array}{lcllcl}
m_c &=& 1.84~~ {\rm GeV}~~, & m_b &=& 5.18 ~~{\rm GeV},\\
\kappa &=& 0.52 ~~{\rm GeV}~~, & a &=& 2.34 ~~{\rm GeV}.
\label{conl}
\end{array}
\end{equation}
The second one was the logarithmic potential\cite{31} given by
\begin{eqnarray}
V(r)&=&-0.6635\;{\rm GeV} +(0.733 \;{\rm GeV}) \log (r\cdot 1 {\rm GeV}),
\end{eqnarray}
with
\begin{equation}
\begin{array}{lcllcl}
 m_c &=& 1.5 ~~{\rm GeV}~~, & m_b &=& 4.906 ~~{\rm GeV}.
\end{array}
\end{equation}
The third one was the improved QCD-motivated potential\cite{8} with the form
\begin{eqnarray}
V(r) &=& \displaystyle \frac{r}{a^2}-\frac{16\pi}{25}\frac{1}{rf(r)}
 \left[\;1+\frac{2\gamma_E+\frac{53}{75}}{f(r)}
 -\frac{462}{625}\frac{\ln f(r)}{f(r)}\;\right],
\end{eqnarray}
where $f(r)$ was given by
\begin{eqnarray}
f(r) &=& 2\ln\displaystyle\left(
     \frac
        {\Lambda_{\overline{\rm MS}} +
         (\Lambda_{\rm II}-\Lambda_{\overline{\rm MS}})
           \exp\left[-\left[15\left( 0.75 \Lambda_{\rm II}
           - \Lambda_{\overline{\rm MS}} \right)r\right]^2\right]
        }
        {\Lambda_{\rm II} \Lambda_{\overline{\rm MS}} \,r
        }
     +C  \right)
\end{eqnarray}
with
\begin{equation}
\begin{array}{lcllcl}
 m_c &=& 1.478 ~~{\rm GeV}~~, & m_b &=& 4.878 ~~{\rm GeV}, \\
 \Lambda_{\rm II} &=& 0.72 ~~{\rm GeV},  & a &=& 2.59 ~~{\rm GeV}^{-1}, \\
 C &=& 4.62. &&&
\end{array}
\end{equation}
To proceed with the calculation we also required an expression for the
running coupling constant $\alpha_s(q)$. The familiar RGE, one-loop result is
\begin{eqnarray}
\alpha_s(q) &=&
\displaystyle\frac{4\pi}{b_0\ln\displaystyle
    \frac{q^2}{\Lambda^2_{\overline{\rm MS}}}},
\label{al1}
\end{eqnarray}
where $b_0=11N_c-2N_f$ and $N_f$ is the number of quark flavors. It is clear
from  Eq.~(\ref{al1}) that $\alpha_s(q)$ contains a Landau singularity in
the nonperturbative region when $q^2=\Lambda^2_{\overline{\rm MS}}$ and
becomes negative for $q^2<\Lambda^2_{\overline{\rm MS}}$. To avoid the
resulting numerical ambiguities we first moved this singularity to $q^2=0$
and used a modified form of $\alpha_s(q)$ in the actual numerical
calculations; namely,
\begin{eqnarray}
\alpha_s(q) &=&
\displaystyle\frac{4\pi}{b_0\ln\left(\displaystyle
 \frac{q^2}{\Lambda^2_{\overline{\rm MS}}}+1\right)}\,.
\label{al2}
\end{eqnarray}
In the next section we shall discuss alternative approaches, the
sensitivity of the results, and their implications. The value of
$\Lambda_{\overline{\rm MS}}$ was taken to be $200$ MeV and $250$ MeV
in the numerical calculations, which is within the experimental range
$\Lambda_{\overline{\rm MS}}=195 + 65 -50 $ MeV\cite{19}. Our numerical
results for the three potentials are presented in Tables I, II, III,
respectively. For comparison we have also included the results for the
$2S$ and $2P$ states. The main features of these results can be summarized
as follows:

\begin{itemize}
\item The results are $\mu$-independent, as they must be.
\item The calculated energy difference between the $J/\Psi$ and the $\eta_c$
mesons is quite close to the experimental value for all three potentials.
\item For each of  these three potentials  we predict the energy difference
between $\Psi'$ and $\eta_c'$ to lie within the range $55-80$ MeV.
\item  For the $b\bar{b}$ system  there are significant discrepancies
between the Cornell model, with the parameters given by Eq.~(\ref{conl}),
and the other two models for the S-states. Since the Cornell model, with
these parameters, does not predict the $b\bar{b}$ spectrum very well, the
results calculated in the other two models are probably better predictions
for the energy difference between the $\Upsilon(1S)$ and the $\eta_b$
($35-50$ MeV) and between the $\Upsilon(2S)$ and the $\eta_b'$ (20 MeV).
\item The predicted energy difference between $B_c^*$ and $B_c$ meson
is in the range $40-70$ MeV from all three of these models, which is
consistent with previous results\cite{14}.
\item The calculated value of $\Delta M_p \equiv E(1^3P_J)-E(1^1P_1)$
for the charmonium $1P$ states is in the range of $-4$ to $-6$ MeV, which
has the same sign but is several times larger than the experimental value
of $-0.9\pm 0.2$ MeV\cite{7}. This is not surprising since there are several
other contributions to $\Delta M_p$ which estimates indicate are comparable
in magnitude to the contribution coming from the hyperfine spin-spin
interaction, $H_{ss}$. In fact, it is surprising that the prediction from
only the one-loop spin-dependent potential is quite close to the experimental
data. We discuss these other contributions in the next section.
\end{itemize}

\section{Discussion and Conclusions}

We have calculated the hyperfine spin splittings in the $c\bar{c}$,
$b\bar{b}$, and $b\bar{c}$ system  using the RGE improved perturbative
spin-spin potential\cite{0}. The results for the hyperfine splittings
of the $S$-wave states agree with the $J/\Psi-\eta_c$ measured
splitting\cite{19} and the prediction for splitting $\Upsilon-\eta_b$ is
reasonable. However, the contribution to $\Delta M_p \equiv
E(^3P_J)-E(^1P_1)$  for the charmonium $P$-wave states is somewhat
larger than the experimental data\cite{7}, although it agrees in sign.
That is, after summing up the leading logarithmic terms and including
them in the perturbation calculations, the agreement with the data is not
as good as the one-loop calculations\cite{8,9,10,11}. In order to illustrate
this  clearly, we can expand $\alpha_s(q)$ in terms of $\alpha_s(\mu)$ and
truncate it at  some finite order. In our final formula, Eq.~(\ref{hssq}),
we used the expansion for $\alpha_s(q)$,
\begin{eqnarray}
\alpha_s(q) &=&
\displaystyle\frac{\alpha_s(\mu)}{1-\displaystyle\frac{b_0}{4\pi}\alpha_s(\mu)
        \ln\frac{\mu^2}{q^2} }
=\alpha_s(\mu)\left[\;1+\sum_{m=1}^{n}\left(\frac{b_0}{4\pi}
 \alpha_s(\mu)\ln\displaystyle \frac{\mu^2}{q^2}\right)^m \;\right],
\label{al3}
\end{eqnarray}
and truncated at several choices of  $n$. Specifically, we repeated the
numerical calculations for the improved QCD motivated potential\cite{8}
for $n=1$, $2$ and $4$, choosing the scale $\mu$, now to be $\mu=1.5$ GeV,
4.0 GeV, and 2.5 GeV for the $c\bar{c}$, $b\bar{b}$, $b\bar{c}$ systems,
respectively. The numerical results are presented in Tables IV, V, and VI
corresponding to $n=1$, $2$, and $4$, respectively. For comparison, we also
presented the results obtained using the complete one-loop hyperfine
potential\cite{4a,4b} in Table VII. Comparing Table IV and Table VII
we see that $\Delta{M}_P$ for $n=1$ is quite close to the complete one-loop
result. However, from Table V and Table VI we see that the predicted values
of $\Delta M_p$ are about 60\%-80\% and 150\%-200\% larger than when terms
up to order 2 and order 4 are kept in the expansion of $\alpha_s(q)$,
Eq.~(\ref{al3}). We note that we also repeated these calculations for
the logarithmic potential\cite{31} and the Cornell potential\cite{30}.
All three potentials predicted similar values for $\Delta M_p$. This
clearly indicates that the nonlocal logarithmic terms from high loop
perturbative calculations are quite important. In fact, even using the
RGE to sum up these logarithmic terms does not allow one to understand
the experimental value of $\Delta M_p$, indicating that the success of
the one-loop calculations\cite{8,9,10,11} was probably fortuitous. In
fact, there are several additional  contributions that are  possibly
comparable in magnitude. These include the following:

The contributions of the spin-orbit and and tensor potentials in
the second order of perturbation theory: These contributions to
$\Delta M_p$ only cancel to first order in perturbation theory.
However, according to the power counting rules introduced in
Ref.\cite{15}, the spin-orbit and tensor potential potentials
shift the energies of the $P$-wave states by an amount of order
$mv^4$ in first order, which indeed cancel in $\Delta M_p$, but they
do make a contribution to $\Delta M_p$ of order $mv^6$ in the second
order of perturbation theory. This estimate is several MeV for the
$P$-wave charmonium states, and therefore should not be neglected.

Higher dimensional operators: Unlike the dimension-six operators,
these give non-zero contributions to $\Delta M_p$ even at tree level.
Compared to the one-loop contribution, these are suppressed by $v^2$ but
enhanced by $\alpha_s^{-1}$ and $v^2/\alpha_s\sim 1$ in charmonium.

The color-octet $S$-wave component in $P$-wave quarkonia states\cite{15}:
This component of the wavefunction receives a tree-level contribution from
the local term $\delta^3({\rm\bf r})$ in the spin-spin potential. This
contribution too could be of order $v^2/\alpha_s\sim 1$ compared to what
has been calculated.

Next-to-leading order perturbative contributions from the two-loop
potential: These are suppressed by order $\alpha_s$, but since $\alpha_s$
is not a very small quantity in  charmonium, one cannot dismiss the
possibility that this contribution could be significant.

Before comparing with the experimental value of $\Delta M_p$ in charmonium,
which is only about $1$ MeV, all the above contribution should be included
since they are possibly comparable in magnitude. In the $b\bar{b}$ case
these effects are less important and one can expect the perturbative
calculations the $b\bar{b}$ system to be more reliable, although
$\Delta M_p$ is smaller, also.

Finally, to explore the sensitivity of our results to the location of the
Landau singularity in $\alpha_s(q)$ we replaced the expression, Eq.~(20), by
\begin{eqnarray}
\alpha_s(q) &=&
\displaystyle\frac{4\pi}{b_0\ln\left(\displaystyle
\frac{q^2}{\Lambda^2_{\overline{\rm MS}}}+\lambda^2 \right)}\,.
\label{al4}
\end{eqnarray}
and varied $\lambda^2$. The results for the $S$- wave hyperfine splitting
were not sensitive to $\lambda^2$ and only for large $\lambda^2$ did
$\Delta M_p$ significantly decrease. To fit $\Delta M_p$ to the measured
value required $\lambda^2$ quite large, about 16, clearly out of the
perturbative region.

We thank Eric Braaten, Yu-Ping Kuang for their instructive discussions
and comments. This work was supported in part by U.S. Department of Energy,
Division of High Energy Physics, under Grant No. DE-FG02-91-ER4086.

\newpage

  \parbox{5.8in}{  Table  I. The hyperfine spin splittings in MeV predicted by
   Eq~(\ref{hssq}) with
  Cornell potential\cite{30}                }

  \begin{tabular}{lrrrrrr} \hline\hline
  & \multicolumn{2}{c}{~~~~$c\bar{c}$      }
  & \multicolumn{2}{c}{~~~~$b\bar{b}$      }
  & \multicolumn{2}{c}{~~~~$b\bar{c}$      }\\
  ~~~~$\Lambda_{\overline{\rm MS}}$ (MeV)
 & ~~~~~~  200~ & ~~~~~~  250~
 & ~~~~~~  200~ & ~~~~~~  250~
 & ~~~~~~  200~ & ~~~~~~  250~
\\ \hline
 $E(1^3S_1)-E(1^1S_0)$~ & 117.1 & 128.3 &  97.7 & 104.3 &  67.0 &  71.4~\\
 $E(2^3S_1)-E(2^1S_0)$~ &  75.3 &  82.5 &  39.6 &  42.2 &  37.7 &  40.3~\\
 $E(1^3P_J)-E(1^1P_1)$~ &  -4.8 &  -7.0 &  -3.0 &  -4.1 &  -4.0 &  -5.6~\\
 $E(2^3P_J)-E(2^1P_1)$~ &  -3.6 &  -5.2 &  -2.1 &  -2.9 &  -3.0 &  -4.2~\\
  \hline \hline \end{tabular} \bigskip \bigskip

 \parbox{5.8in}{ Table  II. The hyperfine spin splittings  in MeV predicted by
   Eq~(\ref{hssq}) with
  Logarithmic potential\cite{31}            }

  \begin{tabular}{lrrrrrr} \hline\hline
  & \multicolumn{2}{c}{~~~~$c\bar{c}$      }
  & \multicolumn{2}{c}{~~~~$b\bar{b}$      }
  & \multicolumn{2}{c}{~~~~$b\bar{c}$      }\\
  ~~~~$\Lambda_{\overline{\rm MS}}$ (MeV)
 & ~~~~~~  200~ & ~~~~~~  250~
 & ~~~~~~  200~ & ~~~~~~  250~
 & ~~~~~~  200~ & ~~~~~~  250~
\\ \hline
 $E(1^3S_1)-E(1^1S_0)$~ & 106.1 & 117.1 &  36.0 &  38.2 &  40.6 &  42.6~\\
 $E(2^3S_1)-E(2^1S_0)$~ &  54.2 &  59.7 &  18.7 &  19.8 &  21.2 &  22.3~\\
 $E(1^3P_J)-E(1^1P_1)$~ &  -5.4 &  -7.8 &  -3.4 &  -4.5 &  -4.5 &  -6.4~\\
 $E(2^3P_J)-E(2^1P_1)$~ &  -2.6 &  -3.8 &  -2.1 &  -2.8 &  -2.8 &  -4.0~\\
  \hline \hline \end{tabular} \bigskip \bigskip

\parbox{5.8in}{  Table  III. The hyperfine spin splittings  in MeV predicted by
   Eq~(\ref{hssq}) with
  Improved-QCD motivated potential\cite{8}  }

  \begin{tabular}{lrrrrrr} \hline\hline
  & \multicolumn{2}{c}{~~~~$c\bar{c}$      }
  & \multicolumn{2}{c}{~~~~$b\bar{b}$      }
  & \multicolumn{2}{c}{~~~~$b\bar{c}$      }\\
  ~~~~$\Lambda_{\overline{\rm MS}}$ (MeV)
 & ~~~~~~  200~ & ~~~~~~  250~
 & ~~~~~~  200~ & ~~~~~~  250~
 & ~~~~~~  200~ & ~~~~~~  250~
\\ \hline
 $E(1^3S_1)-E(1^1S_0)$~ & 107.9 & 119.1 &  44.6 &  47.6 &  43.4 &  45.7~\\
 $E(2^3S_1)-E(2^1S_0)$~ &  68.5 &  75.6 &  20.9 &  22.4 &  25.2 &  26.7~\\
 $E(1^3P_J)-E(1^1P_1)$~ &  -4.6 &  -6.7 &  -2.7 &  -3.7 &  -3.7 &  -5.3~\\
 $E(2^3P_J)-E(2^1P_1)$~ &  -3.4 &  -5.0 &  -1.8 &  -2.5 &  -2.6 &  -3.8~\\
  \hline \hline \end{tabular} \bigskip \bigskip

\eject
\parbox{5.8in}{ Table  IV. The hyperfine spin splittings  in MeV predicted by
   Eq~(\ref{hssq}) with $n=1$ for the
  Improved-QCD motivated potential\cite{8}  }

  \begin{tabular}{lrrrrrr} \hline\hline
  & \multicolumn{2}{c}{~~~~$c\bar{c}$      }
  & \multicolumn{2}{c}{~~~~$b\bar{b}$      }
  & \multicolumn{2}{c}{~~~~$b\bar{c}$      }\\
  ~~~~$\Lambda_{\overline{\rm MS}}$ (MeV)
 & ~~~~~~  200~ & ~~~~~~  250~
 & ~~~~~~  200~ & ~~~~~~  250~
 & ~~~~~~  200~ & ~~~~~~  250~
\\ \hline
 $E(1^3S_1)-E(1^1S_0)$~ & 111.8 & 125.4 &  45.7 &  49.2 &  45.2 &  48.6~\\
 $E(2^3S_1)-E(2^1S_0)$~ &  71.3 &  80.1 &  21.5 &  23.2 &  26.3 &  28.4~\\
 $E(1^3P_J)-E(1^1P_1)$~ &  -1.5 &  -1.9 &  -1.0 &  -1.2 &  -1.2 &  -1.5~\\
 $E(2^3P_J)-E(2^1P_1)$~ &  -1.0 &  -1.3 &  -0.7 &  -0.8 &  -0.9 &  -1.2~\\
  \hline \hline \end{tabular} \bigskip \bigskip

\parbox{5.8in}{ Table  V. The hyperfine spin splittings   in MeV predicted by
   Eq~(\ref{hssq}) with $n=2$ for the
  Improved-QCD motivated potential\cite{8}  }

  \begin{tabular}{lrrrrrr} \hline\hline
  & \multicolumn{2}{c}{~~~~$c\bar{c}$      }
  & \multicolumn{2}{c}{~~~~$b\bar{b}$      }
  & \multicolumn{2}{c}{~~~~$b\bar{c}$      }\\
  ~~~~$\Lambda_{\overline{\rm MS}}$ (MeV)
 & ~~~~~~  200~ & ~~~~~~  250~
 & ~~~~~~  200~ & ~~~~~~  250~
 & ~~~~~~  200~ & ~~~~~~  250~
\\ \hline
 $E(1^3S_1)-E(1^1S_0)$~ & 110.5 & 123.4 &  45.2 &  48.5 &  44.7 &  47.8~\\
 $E(2^3S_1)-E(2^1S_0)$~ &  70.6 &  79.1 &  21.3 &  23.0 &  26.1 &  28.2~\\
 $E(1^3P_J)-E(1^1P_1)$~ &  -2.5 &  -3.4 &  -1.6 &  -2.0 &  -2.0 &  -2.7~\\
 $E(2^3P_J)-E(2^1P_1)$~ &  -1.6 &  -2.2 &  -1.0 &  -1.3 &  -1.4 &  -1.8~\\
  \hline \hline \end{tabular} \bigskip \bigskip

 \parbox{5.8in}{ Table  VI. The hyperfine spin splittings   in MeV predicted by
   Eq~(\ref{hssq}) with $n=4$ for the
  Improved-QCD motivated potential\cite{8}  }

  \begin{tabular}{lrrrrrr} \hline\hline
  & \multicolumn{2}{c}{~~~~$c\bar{c}$      }
  & \multicolumn{2}{c}{~~~~$b\bar{b}$      }
  & \multicolumn{2}{c}{~~~~$b\bar{c}$      }\\
  ~~~~$\Lambda_{\overline{\rm MS}}$ (MeV)
 & ~~~~~~  200~ & ~~~~~~  250~
 & ~~~~~~  200~ & ~~~~~~  250~
 & ~~~~~~  200~ & ~~~~~~  250~
\\ \hline
 $E(1^3S_1)-E(1^1S_0)$~ & 108.8 & 120.4 &  44.8 &  47.9 &  43.9 &  46.5~\\
 $E(2^3S_1)-E(2^1S_0)$~ &  69.3 &  76.8 &  21.0 &  22.7 &  25.6 &  27.4~\\
 $E(1^3P_J)-E(1^1P_1)$~ &  -4.0 &  -6.2 &  -2.3 &  -3.0 &  -3.1 &  -4.3~\\
 $E(2^3P_J)-E(2^1P_1)$~ &  -2.8 &  -4.3 &  -1.4 &  -1.9 &  -2.1 &  -2.9~\\
  \hline \hline \end{tabular} \bigskip \bigskip

\eject

\parbox{5.8in}{ Table  VII. The hyperfine spin splittings   in MeV predicted by
   the complete one-loop spin-spin potential\cite{4a,4b} with
  Improved-QCD motivated potential\cite{8}  }

  \begin{tabular}{lrrrrrr} \hline\hline
  & \multicolumn{2}{c}{~~~~$c\bar{c}$      }
  & \multicolumn{2}{c}{~~~~$b\bar{b}$      }
  & \multicolumn{2}{c}{~~~~$b\bar{c}$      }\\
  ~~~~$\Lambda_{\overline{\rm MS}}$ (MeV)
 & ~~~~~~  200~ & ~~~~~~  250~
 & ~~~~~~  200~ & ~~~~~~  250~
 & ~~~~~~  200~ & ~~~~~~  250~
\\ \hline
 $E(1^3S_1)-E(1^1S_0)$~ & 127.7 & 145.7 &  50.1 &  54.3 &  39.2 &  41.5~\\
 $E(2^3S_1)-E(2^1S_0)$~ &  81.5 &  93.3 &  23.4 &  25.6 &  22.6 &  23.9~\\
 $E(1^3P_J)-E(1^1P_1)$~ &  -1.2 &  -1.5 &  -0.6 &  -0.7 &  -0.6 &  -0.7~\\
 $E(2^3P_J)-E(2^1P_1)$~ &  -1.1 &  -1.3 &  -0.5 &  -0.6 &  -0.5 &  -0.6~\\
  \hline \hline \end{tabular} \bigskip \bigskip

\end{document}